# Coal Strength with Dewatering and Coal Seam Gas Depletion

*Jimmy Xuekai Li[A], Thomas Flottmann[B], Max Millen[B], Shuai Chen[A], Yixiao Huang[A], Zhongwei Chen[A*]*

[A]The University of Queensland, St Lucia, QLD 4072, Australia

[B]Origin Energy, 180 Ann Street, Brisbane, QLD 4000, Australia

*Correspondence to. Email: zhongwei.chen@uq.edu.au

**Abstract.** Understanding the response of coal mechanical properties to dewatering and gas depletion is critical for estimating borehole stability and designing infill coal seam gas (CSG) wells. Despite its importance, the full impact of these processes on coal strength remains little explored. This study aims to quantify these effects through a combination of results from micro-CT imaging, sonic testing, and mechanical testing on coal samples. Micro-CT imaging provides insights into coal's internal structure by focusing on parameters such as fracture porosity and fracture intensity (P32 factor). Sonic testing measures dynamic properties, including P-wave, S-wave velocities ($Vp$ and $Vs$) and dynamic Young's modulus ($Ed$), under both dry and wet conditions. Mechanical testing with acoustic emission (AE) monitoring evaluates static properties like Young's modulus ($Es$) and uniaxial compressive strength (UCS). The key findings are: (i) Micro-CT imaging shows a strong correlation between coal fracture porosity and P32, offering detailed insights into the coal micro-structure; (ii) mechanical testing reveals that dry samples exhibit a 10% higher $Es$ and 31% greater UCS than wet samples, suggesting that dewatering increases coal strength but potentially also promotes embrittlement; and (iii) wet samples show higher $Vp$ and $Ed$ in sonic tests, indicating water saturation significantly influences sonic measurements. These findings improve the understanding of dewatering and gas depletion effects, laying the groundwork for more advanced geomechanical models in coal seam gas (CSG) operations.

# 1. Introduction

Coal seam gas (CSG) extraction is a critical process for meeting global energy demands, particularly as a cleaner alternative to traditional fossil fuels ([Randall 2014](); [Maurin 2017]()). The extraction of CSG involves the simultaneous dewatering of coal seams and depletion of gas, processes that may significantly alter the geomechanical properties of the coal matrix. Understanding these changes is vital for ensuring borehole stability and optimizing the design of infill wells, which are often required to enhance recovery from existing reservoirs ([Askarimarnini 2017](); [Mohanty 2019](); [Zhu et al. 2022]()).

Borehole instability is a common challenge in CSG operations, often exacerbated by the complex interplay of mechanical, hydraulic, and thermal stresses induced by dewatering and gas depletion ([Mohanty 2019]()). These processes influence coal's internal structure, strength, and deformability, leading to potential risks such as wellbore collapse, casing failure, or fines/sand production. Consequently, accurate characterization of coal mechanical properties under varying conditions is essential for safe and efficient drilling operations ([Xie et al. 2019](); [Li et al. 2020]()).

Despite the significance of this issue, a comprehensive understanding of how dewatering and gas depletion impact coal strength and elastic properties remains limited. Existing studies have primarily focused on individual aspects, such as permeability changes or gas migration, without fully integrating the dynamic and static mechanical responses of coal ([Liu et al. 2017](); [Liu et al. 2020]()). Furthermore, the role of microstructural changes, such as fracture porosity and intensity, in influencing these properties is poorly understood ([Li et al. 2020](); [Yu et al. 2020]()). Other prior research has independently investigated aspects such as coal strength ([Poulsen and Adhikary 2013]()), dewatering techniques ([Poulsen et al. 2014]()), and sonic responses ([Pan et al. 2013]()). However, these study pioneers lack the concurrent application of micro-CT imaging, ultrasonic mapping, and mechanical testing to uncover new relationships between microstructural features, such as P32 and fracture porosity, and the mechanical properties of coal, both in dry and wet states.

This study aims to address these gaps by employing an integrated experimental approach that combines micro-CT imaging, sonic testing, and mechanical testing. Micro-CT imaging is used to characterize the microstructural features of coal, including fracture porosity and fracture intensity (e.g., P32 factor). Sonic testing provides dynamic property measurements, such as sonic velocities (i.e., $V_p$ and $V_s$) and dynamic Young's modulus ($E_d$), under both dry and wet conditions. Mechanical testing quantifies static properties, including Young's modulus ($E_s$) and uniaxial compressive strength (UCS), under similar conditions.

The objectives of this study are twofold: (1) to quantify the changes in coal mechanical properties due to dewatering and gas depletion and (2) to establish correlations between static and dynamic properties with structural parameters derived from micro-CT imaging.

# 2. Methodology

Coal samples were sourced from the Bowen Basin, and each sample was cut into cubes with a side of approximately 50 mm using water-cooled diamond saws to minimize thermal damage.

Following preparation, the samples were sent for micro-CT scanning to generate high-resolution 3D digital reconstructions of the coal's internal structure. These digital models captured key microstructural features, such as fractures and minerals, which were used to analyze coal properties.

After CT scanning, the samples were divided into dry and wet groups to simulate different moisture conditions. The dry samples were prepared by drying at room temperature for 24 hours to remove excess moisture. Wet samples were saturated by immersing them in a desiccator filled with distilled water under vacuum conditions for 48 hours to achieve full saturation.

The two groups of samples were further tested for non-destructive ultrasonic mapping and final destructive UCS test with acoustic emission (AE) monitoring. The systematic preparation and testing procedures as illustrated in **Figure 1.**

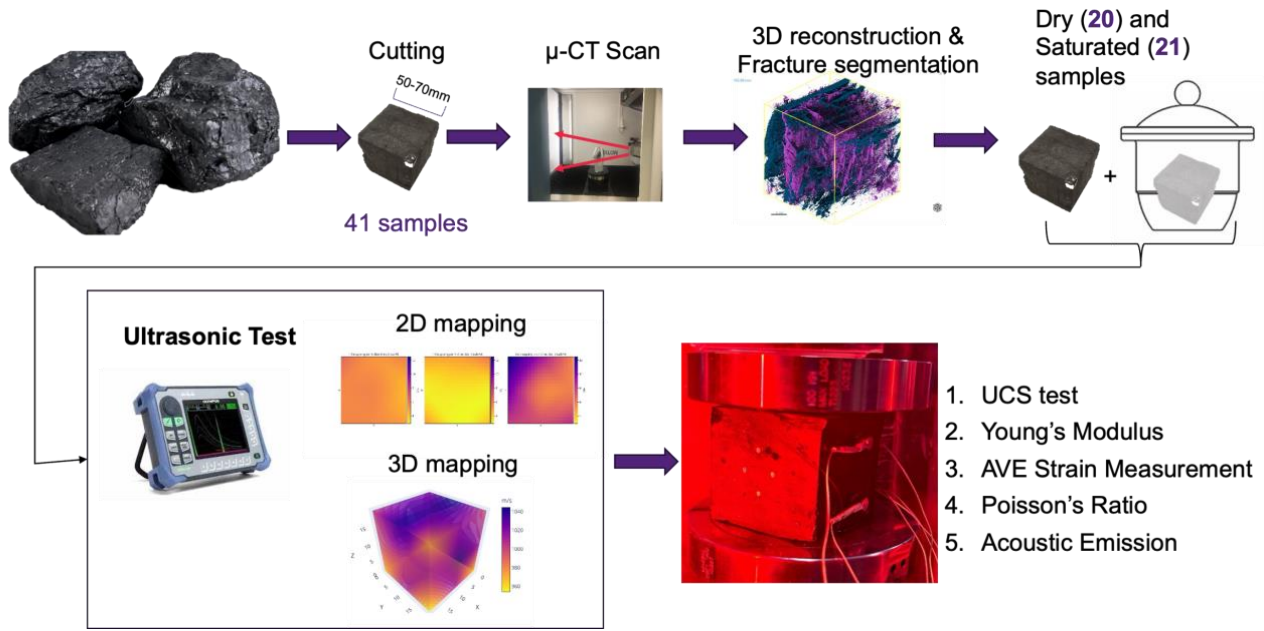

**Figure 1.** Samples preparation and flowchart of the experimental procedures.

## 3. Results

*3.1 Micro-CT Imaging Insights*

Micro-CT imaging was performed using a high-resolution CT scanner (e.g., YXLON Y.MU2000-D) with a spatial resolution of 50 μm to capture detailed imaging of microstructural features.

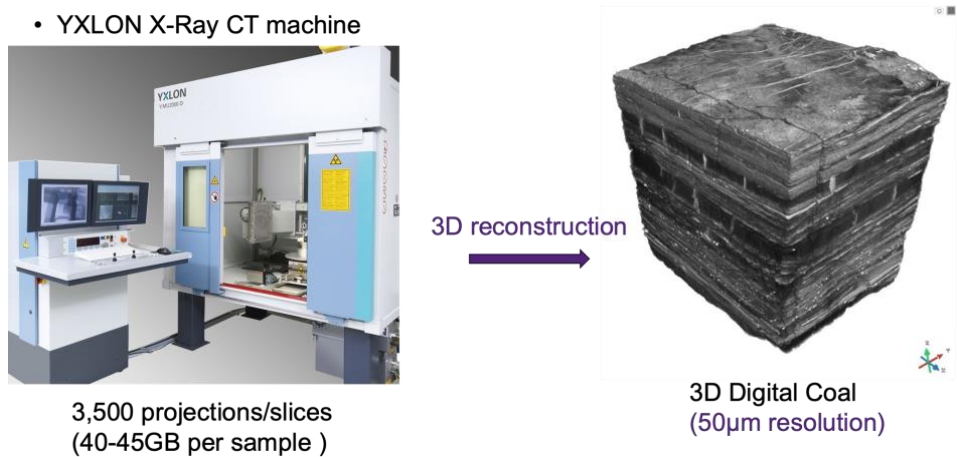

**Figure 2.** The digital coal was reconstructed by using 3,500 projections (40–45 GB per sample) measured via YXLON CT machine.

Each sample was scanned under consistent conditions, capturing 3,500 projections/slices per sample, resulting in approximately 40–45 GB of raw CT imaging data per sample as shown in **Figure 2**. The high-quality 3D digital coal models of the internal structure were created from the CT imaging. The reconstructed images were segmented to accurately differentiate fractures, minerals, and the coal matrix, facilitating detailed analysis of microstructural characteristics (**Figure 3**).

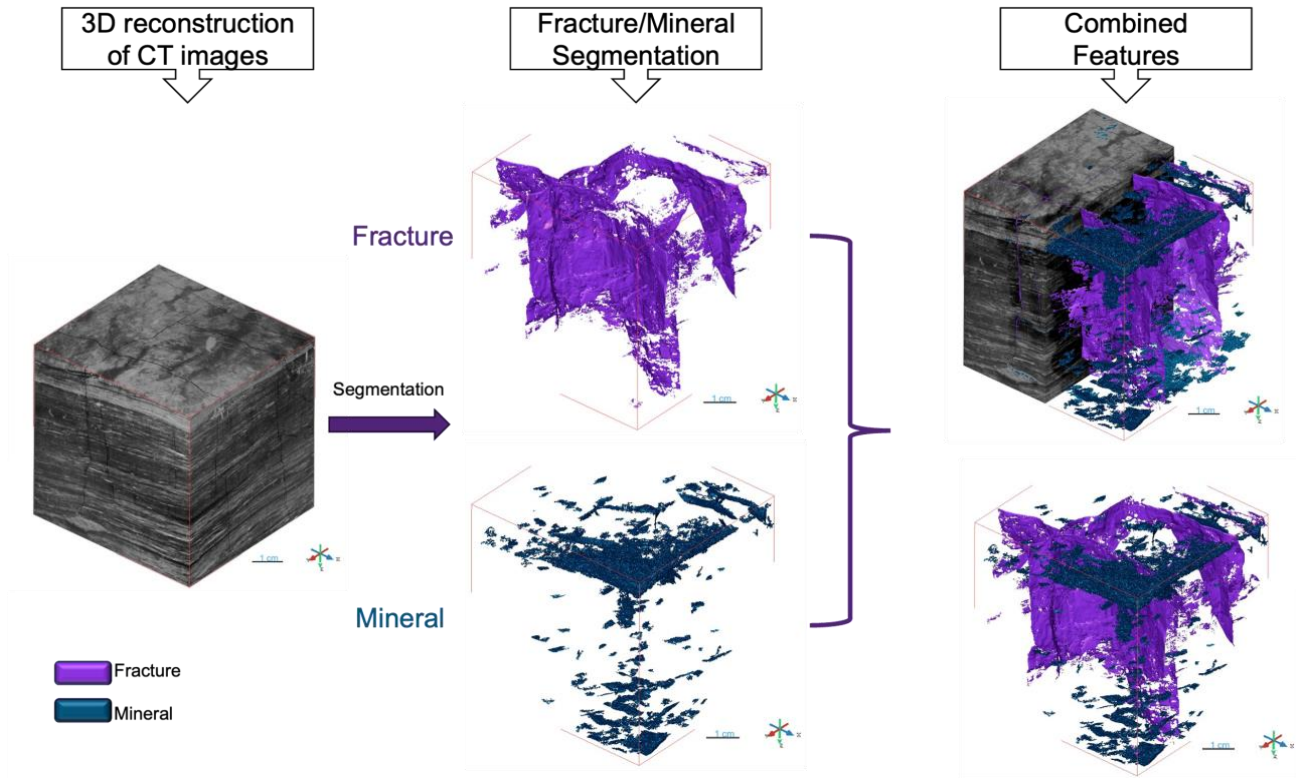

**Figure 3.** High-resolution 3D models of coal samples, with image segmentation applied to distinguish fractures, minerals, and the coal matrix for detailed microstructural analysis.

In addition to reconstructing the 3D digital model of the coal samples, the CT scan analysis provided critical insights into the following parameters:

- **Fracture Porosity**: The volume fraction of fractures within the coal matrix, offering a measure of the void space available for fluid flow or gas storage.

- **Fracture Intensity (P32)**: The total surface area of fractures ($A_t$) per unit volume ($V_u$), indicating the density and distribution of the fracture network within the sample as Eq. (1) (Morelli 2024).

$$P32 = \frac{A_t}{V_u} \tag{1}$$

- **Mineral Content**: The CT imaging also detected and identified mineral inclusions within the coal matrix. These minerals were analyzed for their spatial distribution and potential influence on the coal's mechanical and dynamic behavior.

These parameters were used to evaluate the microstructural characteristics of coal and their relationship with mechanical and dynamic properties, providing a comprehensive understanding of coal internal structure and facilitating the correlation between microstructural features and mechanical behavior.

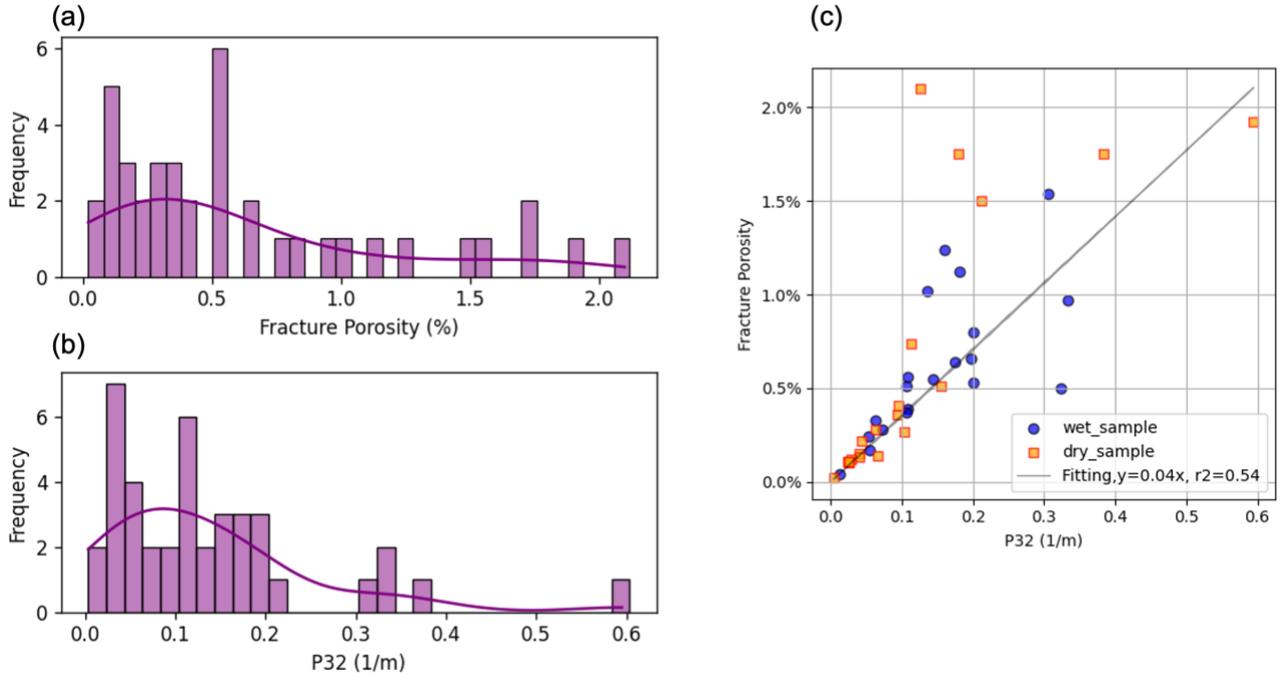

**Figure 4.** *(a)* Histogram illustrating the density distribution of fracture porosity. *(b)* Histogram showing the density distribution of fracture intensity (P32). *(c)* Scatter plot displaying the correlation between fracture porosity and fracture intensity (P32) for both wet and dry samples.

A strong correlation was also observed between fracture intensity (P32) and fracture porosity. Samples with higher fracture porosity consistently showed elevated P32 values, reflecting a denser and more interconnected fracture network (**Figure 4**). These findings highlight the significant influence of microstructural characteristics on the mechanical behavior of coal.

*3.2 Ultrasonic Testing Results*

Ultrasonic testing was conducted using the ultrasonic pulse velocity tester (e.g., Pundit PL-200). Transducers (Olympus V103-RB for P-wave and V153-RB for S-wave measurement) were placed at opposite ends of each coal sample to measure both P- and S-wave velocities.

Each cube sample was marked along the X, Y, and Z axes to define measurement areas, and a non-destructive sonic testing method was used to measure compressional (P-wave, $V_p$) and shear (S-wave, $V_s$) velocities. Dynamic Bulk Modulus ($K$), Dynamic Shear Modulus ($G$) and Dynamic Young's Modulus ($E_d$) are also calculated by using Eq. (2 - 4) ([Gassmann 1951](#)).

$$K = \rho(V_p^2 - \tfrac{4}{3}V_s^2) \quad (2)$$

$$G = \rho V_s^2 \quad (3)$$

$$E_d = \frac{9KG}{3K+G} \quad (4)$$

where,

$K$: Dynamic Bulk Modulus (Pa)

$G$: Dynamic Shear Modulus (Pa)

$E_d$: Dynamic Young's Modulus (Pa)

$\rho$: Density (kg/m$^3$)

$V_p$: P-wave Velocity (m/s)

$V_s$: S-wave Velocity (m/s)

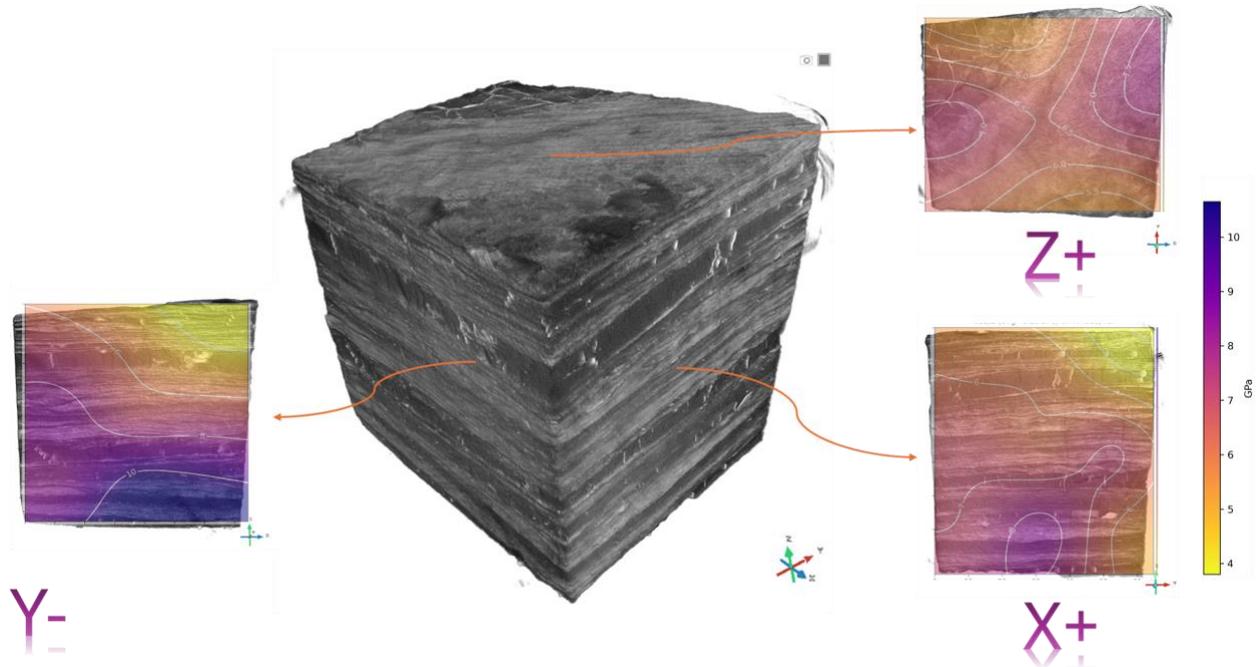

**Figure 5.** The dynamic Young's modulus contour map was developed for each direction of the coal cube.

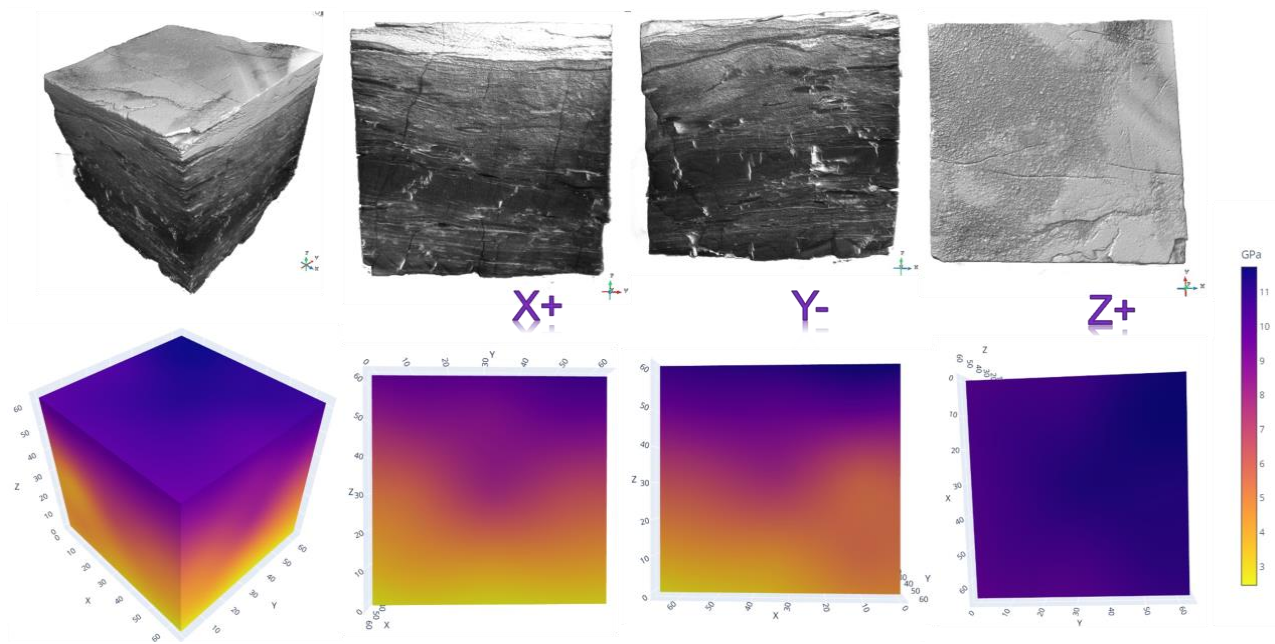

**Figure 6.** Comparison of 2D and 3D ultrasonic mappings (dynamic Young's modulus $E_d$ in GPa) with digital coal models.

Each cube face was divided into 9 equal sections, creating a grid of 27 measurement points per wave type. Velocity data from the X, Y, and Z directions were used to generate 2D ultrasonic maps for each plane, with interpolation algorithms applied to enhance resolution and better represent the coal's properties (

**Figure 5**). Sonic transducers transmitted and received P-wave and S-wave signals at each grid point in all three directions, yielding 27 velocity values for each wave type per sample, capturing the cube's internal structure in 3D (**Figure 6**).

**Figure 7** illustrates the density distributions of dynamic properties, including P-wave velocity (*Vp*), S-wave velocity (*Vs*), dynamic Young's modulus (*Ed*), and dynamic bulk modulus (*K*), for dry and wet coal samples. The dynamic properties (*Vp*, *Vs*, *Ed*, and *K*) are consistently greater in wet samples compared to dry samples. This demonstrates the evident influence of water saturation on the elastic and wave propagation characteristics of coal. The water saturation significantly and noticeably alters the dynamic response of coal, likely due to changes in pore fluid properties and the water bridges of microfractures are created by the pore-filling fluid that enhances the wave propagation (Li *et al.* 2022). Although P-wave velocities (Vp) increased in wet samples due to fluid-enhanced wave propagation, S-wave velocities (Vs) remained less affected, as water does not support shear stresses.

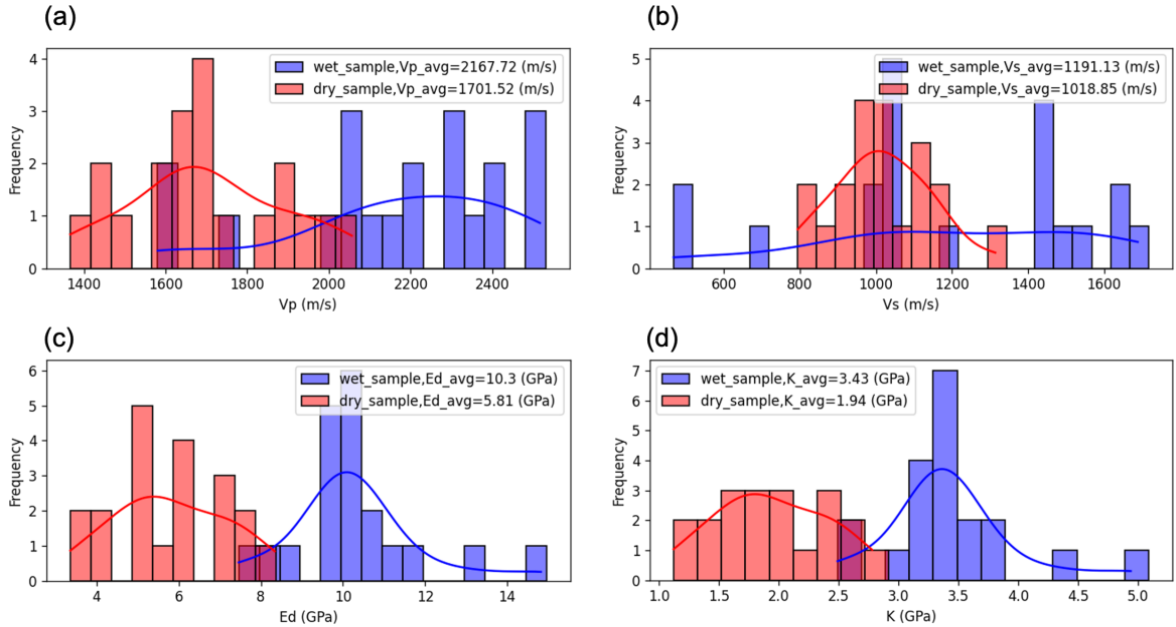

**Figure 7.** Density distributions of dynamic properties for both dry and wet coal samples: *(a)* P-wave velocity (*Vp*), *(b)* S-wave velocity (*Vs*), *(c)* dynamic Young's modulus (*Ed*), and *(d)* dynamic bulk modulus (*K*).

*3.3 Mechanical Testing Results*

Mechanical testing was conducted using a uniaxial compression testing machine (e.g., Instron 5584) to measure the UCS, which is the maximum axial stress sustained by the sample before failure.

The loading rate was maintained at 1 mm/min to ensure quasi-static loading conditions. Advanced non-contacting Video Extensometer (AVE 2) with a cross-polarized lighting system was used to capture axial and lateral deformations for precise measurement of Poisson's Ratio and Young's Modulus, which were determined from the initial linear portion of the stress-strain curve during compression. During mechanical testing, acoustic emissions were monitored using the Mistras Acoustic Emission (AE) monitoring system. The AE count and hit rates and energy were recorded to identify fracture initiation and propagation mechanisms

(**Figure 8**).

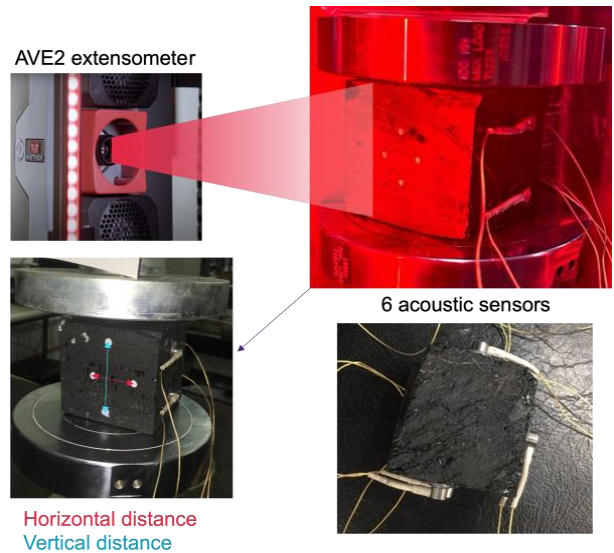

**Figure 8.** The AVE2 extensometer was used for axial and lateral displacement monitoring; The acoustic emission sensors are equipped for cracking (AE events) monitoring during the compression test.

**Figure 9** provides an example of typical stress-strain curves together with AE monitoring data for both dry and wet samples. Correlations between AE activity and structural or mechanical properties provided additional insights into coal failure behavior. The consistency between AE monitoring and the stress-strain curve highlights the strong relationship between the mechanical response of coal and its microstructural evolution under stress. AE data provides real-time insights into the damage mechanisms occurring within the material, complementing the stress-strain analysis.

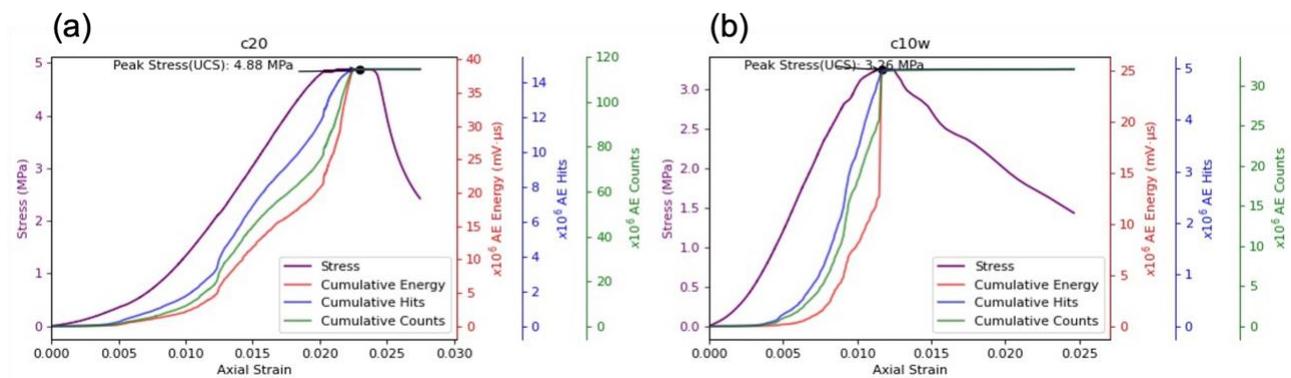

**Figure 9.** Typical stress-strain curves for *(a)* dry sample and *(b)* wet sample.

Separate tests were conducted for dry and wet samples to evaluate the impact of moisture content on static mechanical properties. This comprehensive approach enabled a deeper understanding of the dynamics of the stress-strain progress and mechanical behavior of coal under different moisture conditions.

Mechanical testing revealed notable differences in the static mechanical properties of coal under dry and wet conditions (**Figure 10**). Young's modulus ($E_s$) was found to be higher in dry samples, averaging around 360.48 MPa, which is approximately 10% greater than the average value (327.59 MPa) observed in the wet samples. This reduction in $E_s$ for wet samples suggests that water saturation weakens the coal matrix by diminishing interparticle cohesion. Similarly, UCS was significantly reduced in wet samples, with values averaging 3.61

MPa, compared to the higher average value (4.75 MPa) recorded for dry samples, reflecting a 31% decrease in strength under saturated conditions. These findings underscore the critical role of dewatering in enhancing coal's stiffness and strength, which are essential for maintaining borehole stability during coal seam gas extraction.

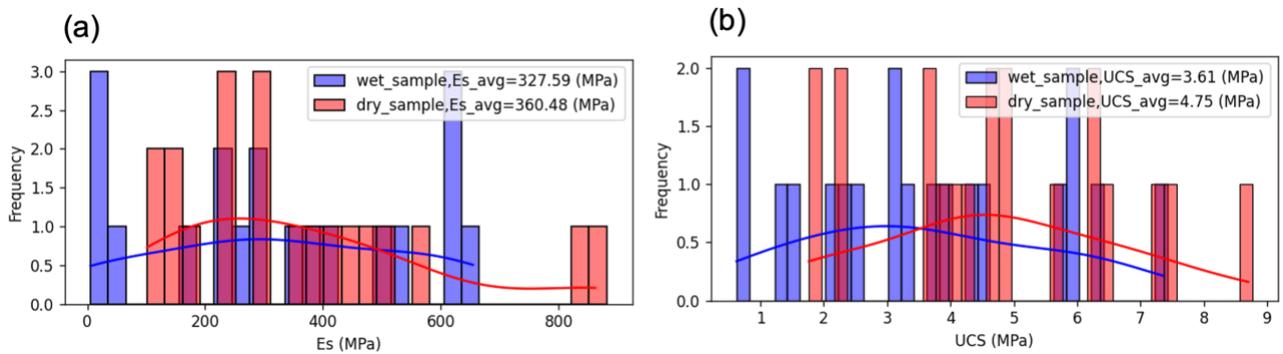

**Figure 10.** Mechanical testing results comparing dry and wet coal samples: *(a)* Young's modulus (Es) shows a 10% higher average value in dry samples (360.48 MPa) compared to wet samples (327.59 MPa); *(b)* uniaxial compressive strength (UCS) is 31% greater in dry samples (4.75 MPa) than in wet samples (3.61 MPa).

The AE monitoring results show that the AE hits and energy are significantly higher in the dry sample compared to the wet sample as shown in **Error! Reference source not found.**. Specifically, the total AE hits in the dry sample exhibit a 104% increase relative to the wet sample, while the total AE energy shows an even more pronounced increase of 340%. This indicates substantially greater acoustic activity in the dry sample during mechanical testing.

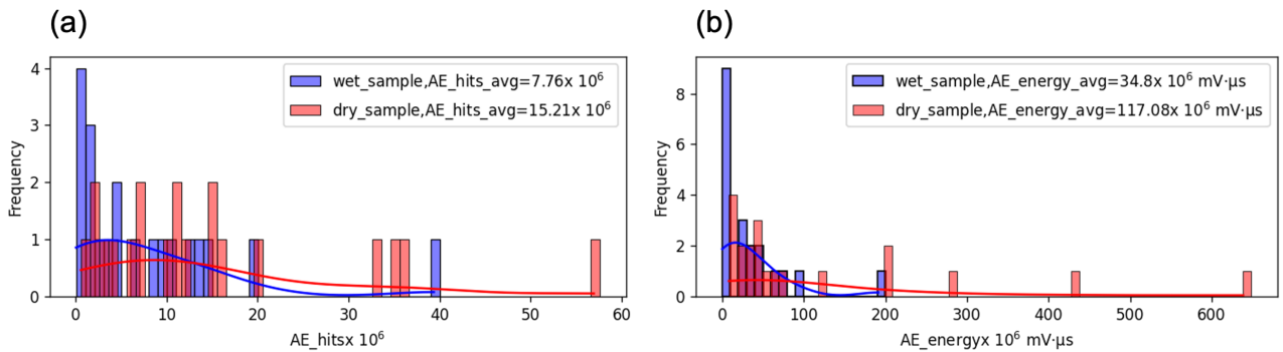

**Figure 11.** *(a)* AE cumulative hits and *(b)* AE cumulative energy demonstrates significantly higher acoustic activity in the dry sample compared to the wet sample, with a 104% increase in total AE hits and a 340% increase in total AE energy.

## 4. Discussion

*4.1 Correlation of Es, UCS and the AE events*

The significantly higher AE hits and energy in dry samples, with a 104% increase in AE hits and a 340% increase in AE energy compared to wet samples (**Error! Reference source not found.**), reflect the impact of moisture content on coal fracture behavior. Our observations are highly consistent with the similar experimental results done by Yu *et al.* (2024). Dry samples, being more brittle, fracture more abruptly, generating greater acoustic emissions. The absence of water in dry samples also removes the damping effect that water provides in wet samples, allowing for more intense AE signals. Moreover, the lack of water in dry samples facilitates the propagation of microcracks, while water in wet samples cushions fractures, suppressing the associated acoustic activity.

These AE observations correlate with the mechanical properties of the coal, particularly *Es* and UCS, which show a direct relationship with AE Hits (**Figure 12**). The dry samples, having a higher average *Es* and UCS, exhibit increased resistance to deformation, leading to greater fracture intensity and more AE events. This aligns with the fact that higher *Es* and UCS generally correlate with greater AE activities, as a material's structural integrity is more stressed before failing. In contrast, the wet samples, with reduced *Es* and UCS, experience less severe deformation and fracture, resulting in lower AE hits and energy.

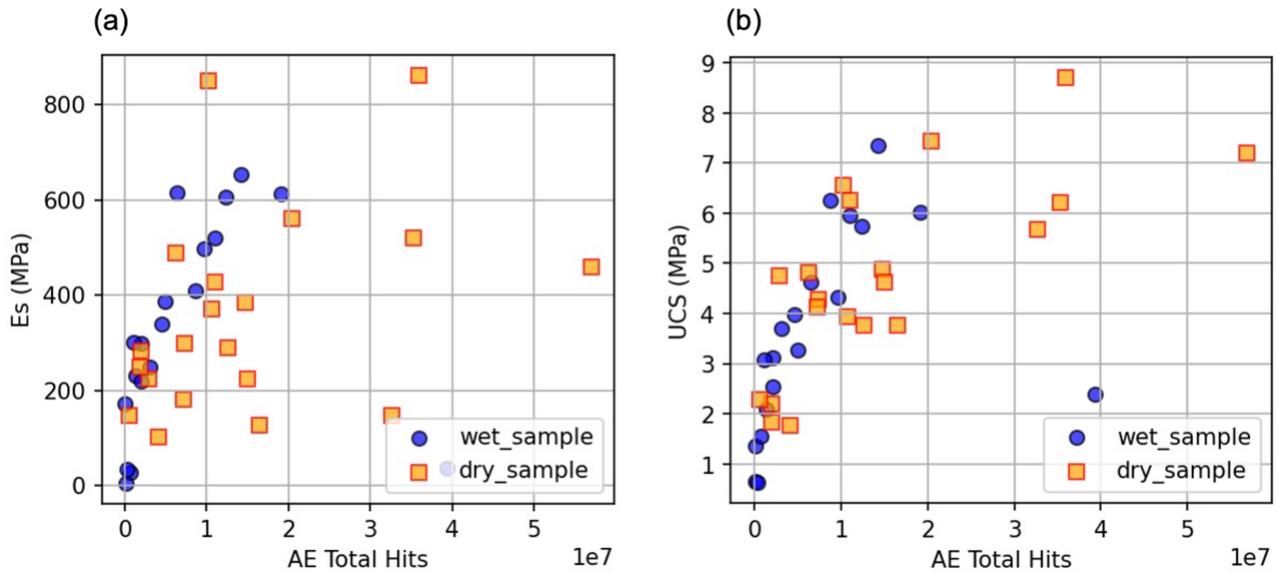

**Figure 12.** *(a)* AE Total Hits v.s. Young's Modulus (*Es*); *(b)* AE Total Hits v.s. UCS.

Together, these findings underscore the significant role that moisture content plays in governing both the mechanical response and fracture behavior of coal. Dry coal, with higher stiffness and strength, generates more AE signals due to more abrupt and extensive fractures, while wet coal, with reduced mechanical strength, shows less AE activity due to the damping effect of water and less intense microcrack propagation.

*4.2 Poisson's Ratio of the Dry and Wet Samples*

Poisson's ratio, which measures the ratio of lateral strain to axial strain under stress, reflects the material's elastic properties and how it deforms when subjected to stress. The observed difference in Poisson's ratio between the wet and dry coal samples is significant. The considerable difference in Poisson's ratio (0.41 for wet samples on average v.s. 0.18 for dry samples on average) further underscores the impact of moisture content on coal mechanical behavior, particularly in terms of elasticity and deformation under stress. The water in the wet samples acts to soften the coal, increasing its lateral strain and overall deformability compared to the drier, stiffer coal.

The higher Poisson's ratio for wet samples indicates that the coal is more compressible and undergoes more lateral deformation when subjected to axial stress. The presence of water likely softens the coal matrix, allowing it to deform more easily in the lateral direction. Water saturation also reduces the material's stiffness by lowering the interparticle cohesion, leading to increased lateral strain for a given axial strain.

In contrast, the lower Poisson's ratio for dry samples suggests that they are less compressible and exhibit less lateral deformation under axial stress. This is consistent with the higher stiffness and brittleness of dry coal, where the absence of water enhances coal's resistance to deformation. Dry coal maintains its shape more rigidly in UCS tests, with less lateral strain for a given axial strain.

*4.3 Limitations of the Study and Future Work*

While micro-CT imaging provides critical insights into coal's microstructure, there are still some limitations needing mitigation in the future work. For example, the spatial resolution of the micro-CT scanner (50 μm) limits the detection of sub-micron fractures and microfracture networks. Fractures narrower than 50 μm, which may contribute significantly to fluid flow and mechanical weakening, are not resolved in the 3D models. This could lead to an underestimation of total fracture porosity and fracture intensity (P32) in highly fractured coals.

The segmentation process, while automated, relies on grayscale thresholds to differentiate fractures, minerals, and the coal matrix. Misclassification can occur where mineral-filled fractures or coal matrix regions exhibit similar X-ray attenuation coefficients, leading to errors in fracture porosity quantification.

CT imaging was conducted under ambient laboratory conditions, neglecting the impact of in situ stresses (e.g., confining pressure) on fracture aperture and connectivity. Fracture closure under reservoir stress could alter porosity and permeability, potentially diverging from the unstressed CT-based measurements.

Moreover, the small specimen size used in this study may not fully represent the heterogeneity of natural coal seams at the reservoir scale. Fracture networks in coal are often scale-dependent, and larger fractures or fracture clusters spanning beyond the specimen volume may influence mechanical behavior in situ. To mitigate the scale effect, multiple samples were used in this study.

Although fracture orientation is recognized as one of key factor influencing sample strength and modulus (Wang *et al.* 2018), there is currently no universally accepted protocol for processing CT data to quantify it. This lack of standardization poses challenges for comparing results across different studies. While future work may incorporate statistical analyses of fracture orientation, this aspect was not the primary focus of the current study.

## 5. Conclusions

This study explored the effects of dewatering and gas depletion on the mechanical properties of coal through an integrated approach combining micro-CT imaging, sonic testing, and mechanical testing. The findings revealed that fracture porosity and fracture intensity (P32) are strongly correlated, offering critical insights into the internal structure of coal and its influence on mechanical behavior. Sonic testing demonstrated that wet samples exhibited higher P-wave velocity ($Vp$) and dynamic Young's modulus ($Ed$), while mechanical testing showed that dry samples had a 10% higher Young's modulus ($Es$) and a 31% greater uniaxial compressive strength (UCS), highlighting the strengthening effect of dewatering. Acoustic emission (AE) analysis further highlighted the impact of moisture content, with dry samples exhibiting significantly higher AE hits (104%) and energy (340%) due to more abrupt fractures and reduced damping effects of water. In contrast, wet samples showed lower AE activity, reflecting suppressed microcrack propagation. These observations align with the strong correlations between microstructural parameters and mechanical properties, underscoring the pivotal role of internal structure and moisture content in coal's mechanical response. Together, these findings enhance the understanding of dewatering and gas depletion effects, laying a foundation for improved geomechanical models in coal seam gas operations.